\documentstyle[amssymb,preprint,aps]{revtex}
\draft

\begin{document}
\title{Unconventional Metallic Magnetism in LaCrSb$_{3}$}
\author{E. Granado$^{1,2}$, H. Martinho$^{3}$, M. S.\ Sercheli$^{3},$ P. G. Pagliuso$%
^{4},$ D.D. Jackson$^{5,6}$, M. Torelli$^{5}$, J. W. Lynn$^{1,2}$, C. Rettori%
$^{3}$, Z. Fisk$^{5}$, and S. B. Oseroff$^{7}$}
\address{$^{1}$NIST\ Center for Neutron Research, National Institute of Standards and%
\\
Technology, Gaithersburg, Maryland 20899 \\
$^{2}$Center for Superconductivity Research, University of Maryland, College%
\\
Park, Maryland 20742\\
$^{3}$Instituto de F\'{i}sica ``Gleb Wataghin'', UNICAMP, 13083-970,\\
Campinas-SP, Brazil\\
$^{4}$Los Alamos National Laboratory, Los Alamos, New Mexico 87545\\
$^{5}$National High Magnetic Field Laboratory, Florida State University,\\
Tallahassee, Florida 32306\\
$^{6}$Lawrence Livermore National Laboratory, Livermore, California 94550\\
$^{7}$San Diego State University,\ San Diego, California 92182}
\date{\today }
\maketitle

\begin{abstract}
Neutron-diffraction measurements in LaCrSb$_{3}$ show a coexistence of
ferromagnetic and antiferromagnetic sublattices below $T_{C}=126$ K, with
ordered moments of 1.65(4) and 0.49(4) $\mu _{B}/$formula unit, respectively
($T$ = 10 K), and a spin reorientation transition at $\approx 95$ K. No
clear peak or step was observed in the specific heat at $T_{C}$. Coexisting
localized and itinerant spins are suggested.
\end{abstract}


The magnetic exchange mechanisms and spin-correlations mediated by itinerant
electrons have attracted renewed interest in the last few years, in part
stimulated by the promising technological possibilities of the emerging
field of spin electronics\cite{Wolf}. The system of intermetallic
antimonides $R$CrSb$_{3}$ ($R=$La-Nd,\ Sm, and\ Gd-Dy) crystallizes in a
layered orthorhombic structure ($Pbcm$ symmetry), with planes of Sb, $R$,
and a layer of CrSb$_{2}$, as shown in Fig. 1(a) \cite{Brylak,Ferguson}. In
this structure, CrSb$_{6}$ octahedra are connected by face-sharing along $%
{\bf c}$, and edge-sharing along ${\bf b}$. The parent compound, LaCrSb$_{3},
$ has been described as an itinerant ferromagnet\cite{Raju,Leonard2,Dubenko}%
.\ The saturation\ magnetization at low-$T$ ($M_{sat})$ depends on the
sample growth procedure, which is typical of itinerant magnets, and values
in the range $0.8\lesssim M_{sat}\lesssim 1.7$ $\mu _{B}/($formula unit,
f.u.) have been reported\cite{Raju,Leonard2,Dubenko,Hartjes,Jackson}. The $R$%
CrSb$_{3}$ compounds with $R=$ Ce, Pr, Nd, and Sm also show a net
ferromagnetic (FM) moment from Cr spins below $T_{C}\sim 105$ K - $147$ K,
with possible ordering of the $R$ moments at much lower $T$\cite
{Leonard2,Hartjes,Leonard1,Deakin}. In contrast, the compounds with $R=$Gd,
Tb, and Dy show no spontaneous FM\ moment at any temperature\cite{Leonard1},
indicating a quantum phase transition in this system.

In this paper we report neutron diffraction and specific heat studies on
crystals of LaCrSb$_{3}$ showing $M_{sat}=1.6\pm 0.1$ $\mu _{B}/$f.u. and $%
T_{C}\sim 126$ K. Neutron diffraction measurements show a coexistence of
orthogonal ferromagnetic (FM) and antiferromagnetic (AFM) spin sublattices.
A spin-reorientation transition in the {\it bc} plane was simultaneously
observed for both magnetic sublattices, by either varying $T$ across $%
T_{sr}\approx 95$ K or by application of small magnetic fields ($H\approx
0.03$ T at $100$ K). Also, no clear peak or step in the specific heat was
observed at $T_{C}$. We suggest a coexistence of localized and itinerant
spins in LaCrSb$_{3}$ to account for these surprising results in a system
showing strong metallic magnetism.

The preparation procedure and characterization of the LaCrSb$_{3}$ samples
used in this work are described in ref. \cite{Jackson}. Specific heat
measurements on single crystals of typical mass $\sim 5-10$ mg were
conducted using the relaxation method, with precision better than $1\%$ and
accuracy better than $10$ \% over the studied $T$-range. dc-magnetization
measurements were conducted using a SQUID magnetometer. Neutron diffraction
experiments using a crystal weighing $\sim 10$ mg were taken in the $(0kl)$
and $(h0l)$ scattering planes, on BT-2, BT-7, and BT-9 triple axis
spectrometers operated in two-axis mode at the NIST\ Center for Neutron
Research. Pyrolytic graphite monochromators and filters were employed using $%
\lambda =2.359(1)$\ \AA\ (BT-2 and BT-9) or $\lambda =2.465(1)$ \AA\ (BT-7),
and the collimations before the monochromator, before and after the sample
were typically $40^{\prime }$ (FWHM). Errors given in parentheses are
statistical only, and represent one standard deviation. The magnetic form
factor of Cr$^{3+}$ was used in the analysis of the magnetic structure\cite
{Brown}. Field-dependent neutron diffraction measurements were performed
using a vertical-field superconducting magnet. A powdered sample was
obtained by grinding small crystallites. Neutron powder diffraction
measurements were carried out with the BT-9 spectrometer and with the BT-1
high-resolution powder diffractometer at NIST. For measurements with BT-1,
monochromatic beams with $\lambda =1.5402(1)$ \AA\ and $2.0783(1)$ \AA\ were
employed. Refinements of the crystal structure of LaCrSb$_{3}$ using powder
data were carried out with the program GSAS using the following values of
the scattering amplitudes: $b($La$)=0.827,$ $b($Cr$)=0.363,$ and $b($Sb$%
)=0.564$ ($\times 10^{-12}$ cm)\cite{Larson}. An orthorhombic model ($Pbcm$
symmetry, see Fig. 1(a)) was employed, with\ good agreements between
observed and calculated intensities at room-$T$. The refined occupancies are 
$1:0.90(3):0.97(2):1.00(2):0.96(2)$ for La, Cr, Sb(1),\ Sb(2), and Sb(3),
respectively. The refined atomic positions and displacement factors at room-$%
T$ do not show significant deviations from those previously reported by
Ferguson {\it et al.}\cite{Ferguson} Room-$T$ $a$, $b$, and $c$ lattice
parameters are $13.2843(5)$ \AA , $6.2119(2)$ \AA , and $6.1191(2)$ \AA ,
respectively. The $T$-dependence of lattice parameters and atomic positions
does not show any observable lattice anomaly or structural transition
between $10$ K and $300$\ K.

Enhancements of some low-angle Bragg reflections due to magnetic ordering
were observed by neutron diffraction at low-$T$. Figure 2(a) shows the $T$%
-dependence of the $(100)$\ reflection (open circles), as well as a fit to a
power law near the transition (solid line), yielding an estimated magnetic
ordering temperature $T_{C}=126.0(6)$ K. The critical exponent is $\beta
=0.34(2)$. Figure 2(b) shows the $T$-dependence of the $(020)$ and $(002)$
Bragg reflections. A magnetic intensity, superposed on the nearly constant
nuclear intensity, was observed for the $(020)$ reflection below $T_{C}$,
and is suppressed below $T_{sr}\approx 95$ K. Conversely, the magnetic
intensity of the $(002)$ reflection is suppressed between $T_{sr}$ and $T_{C}
$ and increases below $T_{sr}.$ The magnetic intensities for the $(100)$, $%
(020)$, and $(002)$ Bragg reflections are due to a FM\ sublattice. Since the
magnetic intensity of a given reflection is sensitive only to the spin
projection into the scattering plane\cite{Bacon}, the suppression of the $%
(002)$ magnetic intensity above $T_{sr}$ (see Fig. 2(b)) shows that the FM\
spins are oriented along the {\it c}-axis for $T_{sr}<T<T_{C}$. Below $%
T_{sr},$ the suppression of the $(020)$\ magnetic intensity shows that the
FM\ moment is oriented along the {\it b}-axis. The FM moment obtained from
the diffraction data at $10$ K is $M_{FM}=1.65(4)$ $\mu _{B}/$f.u.. This is
in very good agreement with $M_{sat}=1.62$ $\mu _{B}/$f.u. obtained from
bulk magnetization measurements on the same crystal.

Interestingly, $(h,0,2l+1)$ Bragg reflections, with nuclear intensities
forbidden by the $Pbcm$ symmetry, were observed below $T_{C}$. Such
reflections, arising from an AFM\ spin sublattice, are relatively weak, and
could be detected only on the single-crystal. Figure 2(c) shows the $T$%
-dependence of the $(001)$ Bragg reflection. Below $T_{sr}$, this reflection
is suppressed, while $(h01)$ Bragg reflections with $h\geq 1$ were still
observed (not shown). This is ascribed to a spin reorientation of the AFM
sublattice to the {\it c}-axis. The ordered AFM moment is $0.49(4)$ $\mu
_{B}/$f.u. at 10 K.

To clarify the relationship between the FM and AFM\ sublattices, neutron
diffraction measurements were performed under the application of a magnetic
field along the {\it b}-axis $(H_{b})$ at 100 K. Figures 2(d), 2(e), and
2(f) show the $H_{b}$-dependence of the nuclear+FM\ $(100)$ and $(002)$ and
AFM $(001)$ Bragg reflections, respectively. The enhancement of the $(002)$
peak and the constant $(100)$ intensity with field $(H_{b}\leq 0.05$ T)
confirms the expected tendency of the FM sublattice to flop from the {\it c}%
-axis to the direction of $H_{b}$. In addition, the suppression of the $(001)
$ reflection for $H\gtrsim 0.03$ T (see Fig. 2(f)), with no significant
increase of the FM moment in the $bc$ plane (see Fig. 2(d)), and the
observation of the (101) AFM\ reflection up to high magnetic fields (7 T,
not shown), shows that the AFM sublattice flops to the {\it c}-axis for a
small field along $b$. The synchronized reorientation of the FM\ and AFM
sublattices, by varying either $T$ or $H_{b}$, reveals that the AFM\ moment
is orthogonal to the FM moment. This result does not support a FM+AFM
inhomogeneous magnetic state for LaCrSb$_{3}$, because in this scenario the
AFM domains are not expected to couple with small magnetic fields\cite
{Wollan}. In contrast, a non-collinear canted spin arrangement trivially
explains the orthogonality between the FM and AFM\ sublattices. The spin
structures that account for the observed magnetic intensities of the (100),
(020), (002) and ($h$01) $(0\leq h\leq 7)$ reflections at 10 K ($\chi
^{2}=1.7)$ and 110 K ($\chi ^{2}=1.1)$ are illustrated in Fig. 1(b). The
total ordered magnetic moment is 1.72(5) $\mu _{B}/$f.u. at $10$ K,
translating into 1.91 $\mu _{B}/$Cr for 90 \% Cr occupancy. The results
above do not exclude the possibility of an inhomogeneous canted state, such
as proposed for lightly Ca-doped LaMnO$_{3}$\cite{Hennion}. Neutron
scattering can be used to investigate the possible inhomogeneities below $%
T_{C}$ and spin correlations above $T_{C}$ in LaCrSb$_{3}$, when larger
crystals become available.

The magnetic state above $T_{C}$ was investigated by means of specific heat
measurements. Figure 3 shows the specific heat, $C(T,H=0)$, and the $dc$%
-magnetization, $M(T,H_{b}=1$ kOe$)$, measured on the same crystal.
Remarkably, neither a peak or step was observed in $C(T_{C})$, within the
resolution of our experiment, which was reproduced for several crystals and
different equipments. To illustrate the significance of this result, the
solid line in Fig. 3(a) shows a simulated magnetic contribution, $%
C_{mag}^{simul}(T)$, with a Gaussian distribution for $C_{mag}^{simul}(T)/T$
with full-width at half-maximum of 20 K and area given by the entropy of the
ideal paramagnetic state for spin-$1$, $S_{PM}=9.1$ J/Kmol. The experimental 
$C(T)$ curve indicates that no measurable change in the magnetic entropy
takes place at $T\approx T_{C},$ despite the relatively sharp magnetic
ordering transition observed by magnetization and neutron-diffraction.
Analyses of $C(T)$ curves commonly yield magnetic entropies above $T_{C}$
smaller than expected for the paramagnetic state, mostly due to magnetic
correlations at $T>T_{C}$ and the difficulty of separating the phononic from
the spin-wave specific heats at $T<T_{C}$. Nonetheless, the absence of any
clear feature in the specific heat of LaCrSb$_{3}$ at $T_{C}$ is striking,
given the large ordered magnetic moments. This result indicates magnetic
correlations above $T_{C}$ which are unusually strong and possibly extended
in range. Such correlations might be low-dimensional in character due to the
layered crystal structure of this compound (see Fig. 1(a)). On the other
hand, the critical exponent below $T_{C}$ obtained from neutron-diffraction, 
$\beta =0.34(2)$ (see Fig. 2(a)), is indicative of three-dimensional spin
fluctuations, at least in the ordered phase.\ The magnetic susceptibility in
the powdered sample (not shown) could be fit by the relation $\chi =\chi
_{0}+C/(T-\Theta _{CW})$ between $\sim 200$ K and 800 K\cite{Raju}, with $%
\chi _{0}=-0.0020$ emu/mol(f.u.), $C=1.13$ Kemu/mol(f.u.), and $\Theta
_{CW}=145$ K. Nonetheless, for a fairly large $T$-interval, between $%
T_{C}=126$ K and $\sim 200$ K, a clear deviation from this behavior was
observed (see also refs. \cite{Raju,Jackson}). This is not surprising in
view of the strong spin correlations above $T_{C}$\ inferred from our
specific heat study.

The coexistence of FM\ and AFM sublattices observed in LaCrSb$_{3}$ is very
unusual in itinerant magnetic systems, which typically show either purely FM
or spin-density-wave AFM spin structures. Nor is it common in insulating
compounds with the spins on transition-metal ions, where the magnetic
properties are described in good approximation by the isotropic Heisenberg
Hamiltonian, and a spiral ground state is the general solution for
equivalent spins\cite{Goodenough}. We should mention that anisotropy terms
in the spin Hamiltonian, originated from spin-orbit coupling, might give
rise to canted spin structures such as those drawn in Fig. 1(b). However, we
emphasize that these terms are typically very small compared to isotropic
exchange for transition-metal compounds, and thus the large angle between
neighboring Cr spins, $36(4)^{\circ }$, is not a result of spin anisotropy.
On the other hand, anisotropy forces must be considered to understand the
abrupt spin reorientation transition at 95 K. This phenomenon is commonly
observed in compounds with competing spin anisotropies from different
magnetic ions, or from the same element at distinct crystallographic sites.
However, in LaCrSb$_{3}$, the Cr ions are placed at the same site. Also, the
spins remain oriented in the same plane at $T_{sr}$ (see Fig. 1(b)), thus
antisymmetric Dzialoshinski-Moriya exchange is not the driving force of this
transition. This indicates at least two species of magnetic electrons in
LaCrSb$_{3}$ with distinct and competing symmetric anisotropies.

We thus suggest a coexistence of localized and itinerant spins in LaCrSb$_{3}
$. The highly anisotropic crystal structure of LaCrSb$_{3}$ may lead to
distinct Cr $3d$ orbitals showing largely different degrees of hybridization
with the Sb states, and therefore different sub-band dispersions. This seems
to be consistent with the spike found close to the Fermi level in the
calculated Cr contribution to the density of states of LaCrSb$_{3}$\cite
{Raju}. As discussed by Korotin {\it et al.} for the possibly similar case
of CrO$_{2}$\cite{Korotin}, the least hybridized sub-band, having almost pure
Cr:$3d$ character, is more sensitive to the on-site Coulomb repulsion, and
tends to become localized. The itinerant spins, from the most hybridized
Cr-(Sb(1),Sb(2)) sub-band, may then polarize the localized spins, leading to
a FM real-charge-transfer exchange mechanism which is closely related to the
double-exchange\cite{Zener}. We note that the electronic transport in LaCrSb$%
_{3}$ is very sensitive to the magnetic order. In fact, the interlayer
resistivity shows a peak at $T_{C}$\cite{Jackson}, while the resistivity
measured along ${\bf b}$ and ${\bf c}$ is similar to calculated curves in
the double-exchange model\cite{Ishizaka}, showing metallic behavior with a
change of concavity at $T_{C}$\cite{Jackson}.

This picture offers insight into the results presented here. At room-$T$,
the Cr-Sb-Cr bonding angles are between 68.3(2)$^{\circ }$ and 68.8(2)$%
^{\circ }$ for neighboring Cr ions along ${\bf c}$, 93.2(2)$^{\circ }$ along 
${\bf b}$, and 133.9(2)$^{\circ }$ along [011] (see Fig. 1), leading to
superexchange interactions of different magnitudes, and possibly different
signs. In addition, a significant Cr-Cr direct bonding\cite{Brylak,Raju} may
give rise to an additional AFM coupling along ${\bf c}$. When such exchange
mechanisms, represented by an isotropic Heisenberg spin Hamiltonian, compete
with real-charge-transfer exchange, a coexistence of FM and AFM sublattices,
such as observed in LaCrSb$_{3},$ can occur. Very recently, Solovyev and
Terakura proposed that either canted spins or FM+AFM phase-separated states
may be stabilized as a result of such a competition in manganites\cite
{Solovyev}. The unusually strong spin correlations in the disordered phase
for LaCrSb$_{3}$, inferred from the specific heat results, suggest that the
localized moments are polarized by the itinerant spins even above $T_{C}$.

In conclusion, we report unconventional metallic magnetism and suggest a
coexistence of localized and itinerant spins in LaCrSb$_{3}$. Whether the
itinerant spins are also responsible for the low-$T$ ordering of the $R$
spin sublattice and the magnetic quantum phase transition in other members
of the $R$CrSb$_{3}$ series\cite{Leonard2,Hartjes,Leonard1,Deakin} is
likely, but confirmation requires more detailed studies.

We thank M. F. Hundley and J. D. Thompson for helpful discussions. This work
was supported by FAPESP and CNPq, Brazil, NSF-MRSEC, DMR 00-80008, NSF DMR
01-02235 and NSF\ DMR 99-71348, USA.

\begin{figure}[tbp]
\caption{(a) Crystal structure of LaCrSb$_{3}$. (b) Projection of the Cr
sublattice onto the {\bf bc} plane. The arrows represent the Cr magnetic
structures in this compound below 95 K ($T_{sr}$) and between $T_{sr}$ and
126 K ($T_{C}$). The angle between neighboring spins, $\alpha =36(4) ^{\circ
}$, is the same within experimental error for both $T$-intervals.}
\end{figure}

\begin{figure}[tbp]
\caption{(a-c) $T$-dependence of the intensities of selected Bragg
reflections: (100) (a), (020) and (002) (b), and (001) (c). The solid line
in (a) is a fit to a power law. The dashed line in (b) is a guide to the
eyes. The vertical dotted lines mark the spin reorientation ($T_{sr}$) and
long-range magnetic ordering ($T_{C}$) transition temperatures. (d-f)
Magnetic-field-dependence of the intensity of (100) (d), (002) (e), and
(001) (f) reflections at 100 K, taken with increasing field (filled
symbols), and with decreasing field after the application of 0.035 T (open
symbols in (f)). The field was applied along {\bf b}.}
\end{figure}

\begin{figure}[tbp]
\caption{$T$-dependence of the $dc$-magnetization (open symbols) and
zero-field specific heat, $C(T)$ (filled symbols), measured for the same
crystal of LaCrSb$_{3}$. The magnetization was measured under an applied
magnetic field of 0.1 T along {\bf b}. The solid line is a simulation for
the magnetic contribution to $C(T)$ (see text).}
\end{figure}

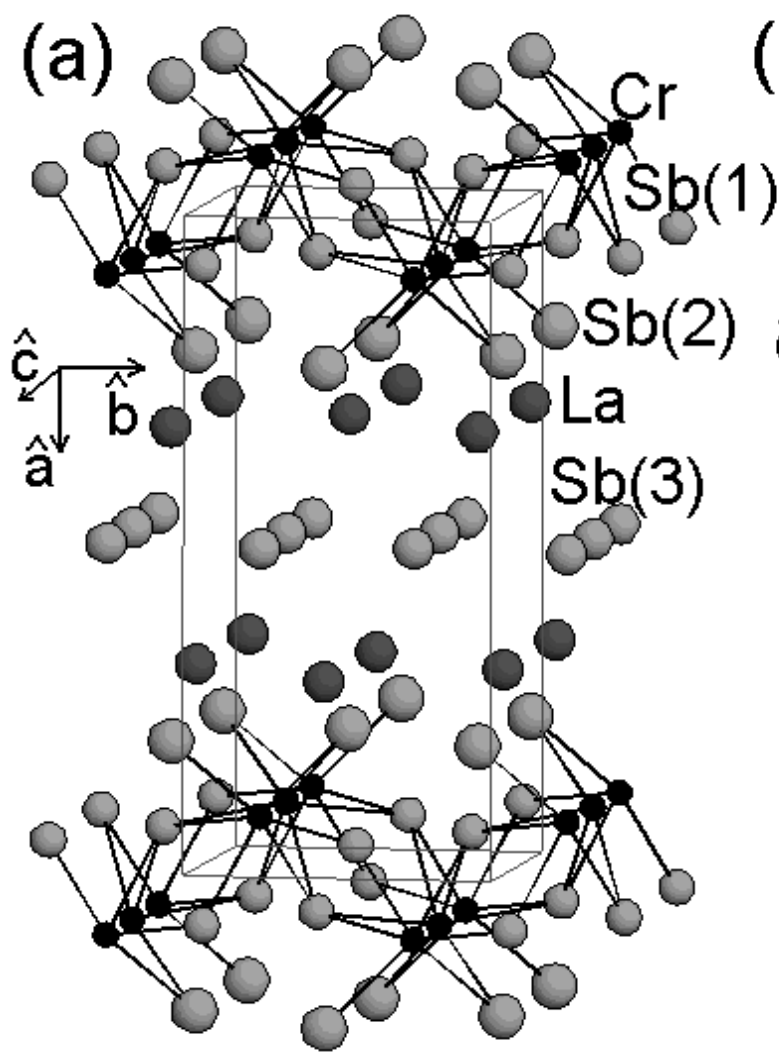
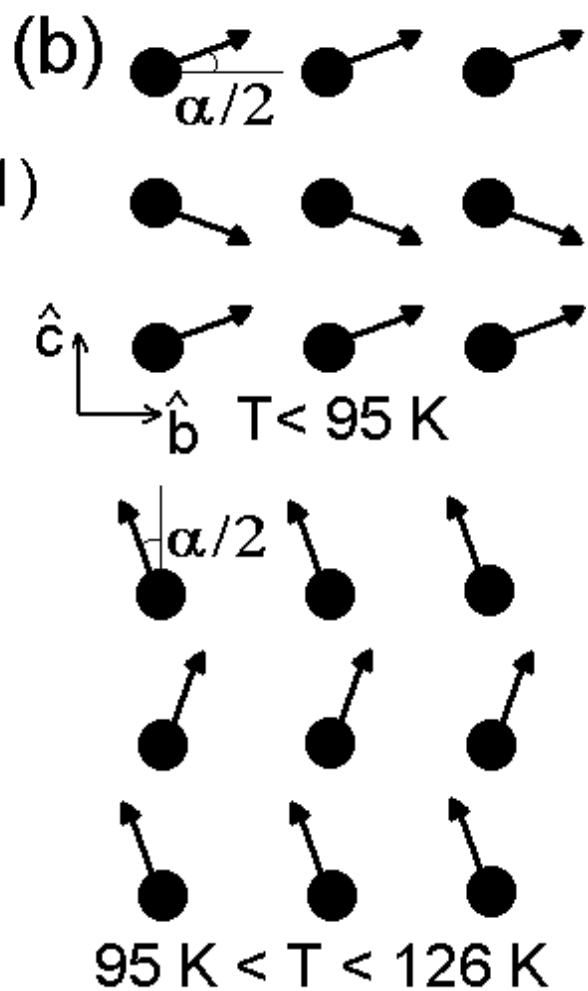
\end{document}